\documentclass[twocolumn,prl,aps,showpacs,floatfix]{revtex4}

\usepackage{graphicx}%
\usepackage{dcolumn}
\usepackage{amsmath}

\newcommand{\lmuh}{\ell_{\mu t}}

\newcommand{\n}{\hspace*{-2.5mm}}

\newcommand{\ice}[1]{\relax}

\newcommand{\re}[1]{(\ref{#1})}

\newcommand{\unl}{\underline}

\newcommand{\als}{\alpha_s}
\newcommand{\beq}{\begin{equation}}

\newcommand{\ba}{\begin{array}}

\newcommand{\ea}{\end{array}}

\newcommand{\eeq}{\end{equation}}

\newcommand{\bea}{\begin{eqnarray}}

\newcommand{\eea}{\end{eqnarray}}

\newcommand{\GeV}{\mbox{\rm GeV}}

\newcommand{\promille}{%
  \relax\ifmmode\promillezeichen
        \else\leavevmode\(\mathsurround=0pt\promillezeichen\)\fi}
\newcommand{\promillezeichen}{%
  \kern-.05em%
  \raise.5ex\hbox{\the\scriptfont0 0}%
  \kern-.15em/\kern-.15em%
  \lower.25ex\hbox{\the\scriptfont0 00}}

\begin{document}

\title{

{
 \vspace*{-14mm}
\centerline{\normalsize\hfill \tt SFB/CPP-06-19}
\centerline{\normalsize\hfill \tt TTP06-14 }
\centerline{\normalsize\hfill \tt hep-ph/0604194}
\baselineskip 11pt
{}
}
\vspace{3mm}

Top Quark Mediated Higgs Decay into Hadrons to Order $\alpha_s^5$
\vskip.3cm
     }
\author{P.~A.~Baikov}
\affiliation{
Skobeltsyn Institute of Nuclear Physics,
        Moscow State University,
        Moscow~~119992, Russia
        }
\author{K.~G.~Chetyrkin}\thanks{{\small Permanent address:
Institute for Nuclear Research, Russian Academy of Sciences,
 Moscow 117312, Russia.\\}}
\affiliation{Institut f\"ur Theoretische Teilchenphysik,
  Universit\"at Karlsruhe, D-76128 Karlsruhe, Germany}

\begin{abstract}
\vspace*{.7cm}
\noindent
We present in analytic form the  ${\cal O}(\alpha_s^5)$ correction
to the $H\to gg$ partial width of the standard-model Higgs boson with
intermediate mass $M_H <  2 M_t$.
Its knowledge is useful  because the ${\cal O}(\alpha_s^4)$ correction is 
sizeable (around  20\%).
For $M_H=120$~GeV, the resulting QCD correction factor reads
$1+(215/12)\alpha_s^{(5)}(M_H)/\pi
+152.5\left(\alpha_s^{(5)}(M_H)/\pi\right)^2
+381.5 \left(\alpha_s^{(5)}(M_H)/\pi\right)^3
\approx1+0.65+0.20 + 0.02$.
The new four-loop correction  increases the total Higgs-boson
hadronic width by a small amount of order 1\,\promille \ and  stabilizes
significantly the residual scale dependence.

\end{abstract}


\maketitle

\section{Introduction}   

Within the Standard Model (SM) the scalar Higgs boson is responsible
for  mechanism of the electroweak mass generation.  It is the last fundamental
particle in the SM which has not yet been directly observed. Its
future \mbox{(non-)discovery} will be of primary importance for all
the particle physics.  The SM Higgs boson mass is constrained from
below, $M_H > 114\,  \GeV$, by experiments at LEP and SLC
\cite{Barate:2003sz,:2004qh}.  Indirect
constraints  from precision electroweak measurements
\cite{Carena:2002es} set an upper limit of 200 $\GeV$ on $M_H$.

Adopting the framework of the SM, the coupling of the  Higgs boson to  
gluons  is mediated by virtual massive quarks \cite{Wilczek:1977zn}
and it is  this coupling which  plays a crucial
r\^ole in Higgs phenomenology.  Indeed, with  the  Yukawa couplings of 
the Higgs boson to  quarks  being proportional to the respective quark
masses, the $ggH$ coupling of the SM is essentially generated by the
top quark alone.  The $ggH$ coupling strength becomes independent of
the top-quark mass $M_t$ in the limit $M_H\ll 2M_t$.  

The process of the gluon-fusion, $gg \to H$, provides a very important
Higgs production mechanism over the all $M_H$ range under
consideration.  The corresponding cross-section is
significantly increased, by approximately 70\%, by 
next-to-leading order (NLO) QCD corrections,  available since long
\cite{Dawson:1990zj,Graudenz:1992pv,Spira:1995rr,Djouadi:1991tk}. 
The largeness of the correction along with the large residual
scheme dependence of the result have motivated the calculations of the
NNLO terms \cite{Harlander:2002wh,Anastasiou:2002yz,Ravindran:2003um}. 
Very recently even the leading N${}^3$LO
corrections to the inclusive cross-section have  been computed \cite{Moch:2005ky}. 
As a  result of these remarkable theoretical advances the theoretical
uncertainty  of the production cross section is reduced significantly 
and is estimated around 20\%. 

The QCD corrections to the closely-related process---the production
cross-section of the Higgs decay into two gluons---are presently
known to NNLO \cite{Chetyrkin:1997iv,Chetyrkin:1997un} only. 
Recently it was pointed out in work \cite{Anastasiou:2005pd} that the
ratio of the production cross-section to the decay rate   is significantly less
(by a factor of 2) sensitive to higher order QCD corrections 
than the individual observables, since 
the corresponding K-factors are similar in size  and tend to cancel 
to a significant  extent. The work also argues that
it is this ratio which presents  the theoretical input  to analyses of Higgs couplings
extractions at the LHC. 
Thus, the knowledge of the N${}^3$LO QCD corrections
to the Higgs decay rate into gluons is highly desirable.

In this publication  we present in   analytic 
form the four-loop ${\cal O}(\alpha_s^5)$ correction
to the $H\to gg$ partial width of the standard-model Higgs boson with
 mass $M_H  < 2 M_t$.

\section{Calculation and Results}

We start by constructing  
an effective Lagrangian, ${\cal L}_{\rm eff}$, by integrating out
the top quark ~\cite{Inami:1982xt,Chetyrkin:1997iv}.
This Lagrangian is a linear combination of certain dimension-four operators
acting in QCD with five quark flavours, while all $M_t$ dependence is
contained in the coefficient functions.
We then renormalize this Lagrangian and compute with its help  the $H\to gg$ decay
width through ${\cal O}(\alpha_s^5)$.

The effective  Lagrangian can be written in the form
\begin{equation}
\label{eff}
{\cal L}_{\rm eff}=-2^{1/4}G_F^{1/2}HC_1\left[O_1^\prime\right].
\end{equation}
Here, $\left[O_1^\prime\right]$ is the renormalized counterpart of the bare
operator $O_1^\prime=G_{a\mu\nu}^{0\prime}G_a^{0\prime\mu\nu}$, where
$G_{a\mu\nu}$ is the colour field strength. The superscript 0 denotes bare
fields, and primed objects refer to the five-flavour effective theory,
$C_1$ is the corresponding renormalized coefficient function, which carries
all $M_t$ dependence.

Eq.~(\ref{eff}) directly leads to  a general expression for the $H\to gg$
decay width,
\begin{equation}
\label{mas}
\Gamma(H\to gg)=\frac{\sqrt2G_F}{M_H}C_1^2
{\rm Im}\, \Pi^{GG}(q^2 = M_H^2)
{},
\end{equation}
where 
\beq
\Pi^{GG}(q^2) = \int \, e^{iqx} \langle 0|
T\left( 
\left[O_1^\prime\right](x)\left[O_1^\prime\right](0)
\right)
|0\rangle
{\rm\, dx}
\label{GGGG}
\eeq
is the vacuum polarization induced by the gluon operator
at $q^2=M_H^2$, with $q$ being the external four-momentum.

It is customary to write eq. \re{mas} in the form
\begin{equation}
\label{mas2}
\Gamma(H\to gg)= K\,  \Gamma_{\rm Born}(H\to gg)
{},
\eeq
where ($G_F$ is Fermi's constant)
\beq
\label{born*K}
\Gamma_{\rm Born}(H\to gg)
=\frac{G_FM_H^3}{36\pi\sqrt2}
\left(\frac{\alpha_s^{(n_l)}(M_H)}{\pi}\right)^2,
\end{equation}
and so-called $K$-factor reads:
\beq
K = \frac{72\, \pi^3}{M_H^4}
\,\, \frac{  C_1^2 \,\, {\rm Im}\,\, \Pi^{GG}(q^2 = M_H^2)}{\left(\alpha_s^{(n_l)}(M_H)\right)^2}
 = 1+\dots
\label{K}
\eeq

The coefficient function  $C_1$ is  known in N${}^3$LO   
\cite{Chetyrkin:1998un,Schroder:2005hy,Chetyrkin:2005ia} 
and reads
\begin{widetext}
\begin{eqnarray}
C_1^{\rm OS}&\n=\n&
-\frac{1}{12}\,\frac{\alpha_s^{(n_f)}(\mu)}{\pi}
\Bigg\{1 
+ \frac{\alpha_s^{(n_f)}(\mu)}{\pi}
\Bigg(
\frac{11}{4} 
- \frac{1}{6} \lmuh
\Bigg)
+ \left(\frac{\alpha_s^{(n_f)}(\mu)}{\pi}\right)^2
\Bigg[
\frac{2693}{288} 
- \frac{25}{48} \lmuh
+ \frac{1}{36} \lmuh^2
+ n_l\left(
-\frac{67}{96} 
+ \frac{1}{3} \lmuh
\right)
\Bigg]
\nonumber\\
&\n\n&{}+ \left(\frac{\alpha_s^{(n_f)}(\mu)}{\pi}\right)^3
\Bigg[
-\frac{4271255}{62208} 
-\frac{2}{3}\zeta(2)\left(1+\frac{\ln2}{3}\right)
+ \frac{1306661}{13824}\zeta(3)
- \frac{4937}{864} \lmuh
+ \frac{385}{144} \lmuh^2
- \frac{1}{216} \lmuh^3
\nonumber\\
&\n\n&{}+  n_l \left(
  \frac{181127}{62208}
+ \frac{1}{9}\zeta(2) 
- \frac{110779}{13824}\zeta(3)
+ \frac{109}{48} \lmuh
+ \frac{53}{96} \lmuh^2
\right) 
+ n_l^2 \left(
- \frac{6865}{31104} 
+ \frac{77}{1728} \lmuh 
- \frac{1}{18} \lmuh^2
\right)
\Bigg]
\Bigg\}
\nonumber\\
&\n\approx\n&
-\frac{1}{12}\,\frac{\alpha_s^{(n_f)}(M_t)}{\pi}
\Bigg[1
+ 2.75000
\frac{\alpha_s^{(n_f)}(M_t)}{\pi}
+  5.86111
\left(\frac{\alpha_s^{(n_f)}(M_t)}{\pi}\right)^2
+  5.39967
\left(\frac{\alpha_s^{(n_f)}(M_t)}{\pi}\right)^3\Bigg]
{}.
\label{eqc1os}
\end{eqnarray}
\end{widetext}

Here $\lmuh = \ln\frac{\mu^2}{M_t^2}$, with $M_t$ being the on-shell
top quark mass; we have displayed for generality the result with the
effective number of light quark flavors denoted as  $n_l$. 
In the
numerical evaluation we have set $\mu =M_t$ and $n_l=n_f-1=5$.
$\zeta_n\equiv \zeta(n)$ is the Riemann's Zeta-function.  

Thus we are left with the calculation of the  last factor in
Eq.~(\ref{mas}), namely the absorptive part $\Pi^{GG}$ in N${}^3$LO, that
is to $\als^3$.  In fact it turned out to be more 
convenient   to calculate the correlator  \re{GGGG} 
per se and take its   absorptive part subsequently.  
Since  $\Pi^{GG}$  starts in  the leading order
from a one-loop diagram, the   ${\cal O}(\alpha_s^3)$ calculation 
faces  as many as 10240 {\em four loop} diagrams
(at NNLO \cite{Chetyrkin:1997iv}  the number was 403).

The overall strategy of our calculations was identical to that used by us
before in works \cite{Baikov:2004tk,Baikov:2005rw}. First we generate the contributing diagrams
with the package QGRAF \cite{Nogueira:1991ex}.
Second, using the criterion of irreducibility of Feynman integrals
\cite{Baikov:criterion:00,Baikov:2005nv}, 
the set of irreducible integrals involved
in the problem was constructed.  Third, the coefficients multiplying
these integrals were calculated as series in the $1/D\rightarrow0$
expansion with the help of an auxiliary integral representation 
\cite{Baikov:tadpoles:96}. Fourth, the exact answer, i.e.  a rational function of $D$,
was reconstructed from this expansion.  The major part of the
calculations was performed on the SGI Altix 3700 computer (32
Itanium-2 1.3 GHz processors) using the parallel version of FORM
\cite{Vermaseren:2000nd,Fliegner:1999jq,Fliegner:2000uy} 
and  took about two weeks  of calendar time in total.

After renormalization, our  result reads
\beq
{\rm Im}\, \Pi^{GG}(q^2) = q^4\,\frac{2}{\pi}\left\{
1+ \sum_{i=1}^{\infty} g_i \,(a_s')^i
\right\}
\label{ImGGGG}
{},
\eeq
\bea
g_1 &=&
 \, n_l 
\left[
-\frac{7}{6}\right]
{+}
 \frac{73}{4}
{},
\label{NImG2G2_1}
\\
{g}_2
&=&  
 n_l^2
\left[
\frac{127}{108} 
-\frac{1}{36}  \,\pi^2
\right]
{+} \, n_l 
\left[
-\frac{7189}{144} 
+\frac{11}{12}  \, \pi^2
+\frac{5}{4}  \,\zeta_{3}
\right]
\nonumber\\
&{+}&
\frac{37631}{96} 
-\frac{121}{16}  \,\pi^2
-\frac{495}{8}  \,\zeta_{3}
{},
\label{NImG2G2_2}
\eea
\bea
{g}_3 &=&   
 n_l^3
\left[
-\frac{7127}{5832} 
+\frac{7}{108}  \,\pi^2
+\frac{1}{27}  \,\zeta_{3}
\right]
\nonumber\\
&{+}& \, n_l^2
\left[
\frac{115207}{1296} 
-\frac{1609}{432}  \,\pi^2
-\frac{113}{24}  \,\zeta_{3}
\right]
\label{NImG2G2_3}
\\
&{+}& \, n_l 
\left[
-\frac{368203}{216} 
+\frac{18761}{288}  \,\pi^2
+\frac{11677}{48}  \,\zeta_{3}
-\frac{95}{36}  \,\zeta_{5}
\right]
\nonumber\\
&{+}&
\left[
\frac{15420961}{1728} 
-352  \,\pi^2
-\frac{44539}{16}  \,\zeta_{3}
+\frac{3465}{8}  \,\zeta_{5}
\right]
{},
\nonumber
\end{eqnarray}
where $a_s'$ stands  for $\alpha_s^{(n_l)}/\pi$.

In eqs.~(\ref{NImG2G2_1}-\ref{NImG2G2_3}) we have set $\mu^2 = q^2$, the full 
$\mu$ dependence can be easily recovered with the standard RG techniques  from
the fact the  product
\[
\left(\beta^{(n_l)}(a'_s)\right)^2 \,  \Pi^{GG}(a'_s,\mu/q^2)
{}
\label{scale} 
\]
is scale independent (\cite{Inami:1982xt}).
In numerical form   ${\rm Im}\, \Pi^{GG}$ reads  (we set  $n_l=5$) 
\bea
 \frac{2\,q^4}{\pi}\,{\rm Im}\, \Pi^{GG} &=& 
\label{ImGG:num}
\\ 
&{}&
\hspace{-2cm}
1 + 12.4167 \, a_s + 68.6482\, a_s^2  - 212.447 \,a_s^3
{}.
\nonumber
\eea

In order to better understand  the structure of the $\alpha_s^3$ term in
\re{ImGG:num} it is  instructive to separate  the genuine four-loop  contributions from
the function $\Pi^{GG}(q^2)$ from  essentially
``kinematical'',  so-called $\pi^2$-terms  originating  from the analytic continuation.
For a given order in
$\alpha_s$ these extra contributions are completely predictable from
the standard evolution equations  applied to the ``more leading'' terms in $\Pi^{GG}(q^2)$
proportional to  some smaller   powers of   $\alpha_s$.
The corresponding expression for ${\rm Im}\, \Pi^{GG}$ assumes the form
\bea
 \frac{2\,q^4}{\pi}\,{\rm Im}\, \Pi^{GG} &=& 1 + 12.4167 \, a_s
\\ 
&{}&
\hspace{-2.3cm}
+\, (104.905 - \unl{36.257})\, a_s^2  +( 886.037 - \unl{1098.48})\,a_s^3
{},
\nonumber
\eea
where we have underlined the contributions coming from analytic
continuation.  Thus, the present calculation confirms the pattern first
observed on the example of the scalar correlator in work \cite{Baikov:2005rw}: 
the ``kinematical'' $\pi^2$ terms tend to neatly cancel
the genuine higher order contributions.

We are now in a position to find the ${\cal O}(\alpha_s^3)$ term of the
$K$ factor in Eq.~(\ref{K}).
To this end, we first multiply $C_1^2$  by $R^{G}(q^2=M_H^2)$, then
eliminate $\alpha_s^{(6)}(\mu)$ in favour of $\alpha_s^{(n_l)}(\mu)$ 
\cite{Bernreuther:1981sg,Chetyrkin:1997sg,Chetyrkin:1998un}
and, finally,  choose the $\mu=M_H$ (to get  a compact expression). The result 
reads:
\begin{widetext}
\beq
K =
1 + 17.9167 \, a'_s +   (156.81 - 5.7083 \ln\frac{M_t^2}{M_H^2}) (a'_s)^2 + 
     (467.68 - 122.44\ln\frac{M_t^2}{M_H^2}  + 10.94 \ln^2\frac{M_t^2}{M_H^2} )\ (a'_s)^3 
{}.
\eeq
\end{widetext}
If we also use $M_t = 175 $ GeV, $M_H =120$ GeV and $\alpha_s^{(5)}(M_H)/\pi = 0.0363$
we arrive at
\bea
K &=& 1 + 17.9167 \, a'_s +   152.5 (a'_s)^2 + 
    381.5 (a'_s)^3 
\nonumber
\\
&=& 1 + 0.65038  + 0.20095   + 0.01825 
{}.
\eea
Thus, we observe that, unlike the NLO and NNLO cases, the ${\cal O}(\alpha_s^3)$
correction,  being by a factor more than   ten less than
the previous $\alpha_s^2$ one,   has quite a moderate  size.
The welcome  stability of the perturbation theory  is also confirmed by 
testing the scale dependence of the $K$-factor. By changing the scale $\mu$ 
from  $M_H/2$ to $2\,M_H$ we find that the maximal deviation of the 
$K$-factor from its value at $\mu =M_H$ decreases from 24\% (LO) to 22\% (NLO), 10\% (NNLO)
and, finally, to only 3\% at NNNLO!

\section{Conclusion}

We have computed the N${}^3$LO  correction of order $\alpha_s^5$  to the
$H\to gg$ partial width of the standard-model Higgs boson with
intermediate mass $M_H <  2 M_t$. Our calculation significantly
reduces the theoretical uncertainty of the QCD prediction 
for this important process.

We are grateful to J.H. K\"uhn for his interest in the work,
numerous discussions and careful reading the manuscript.
We want to thank M. Steinhauser for his  timely reminding the
importance of   the calculation. 

This work was supported by 
the Deutsche Forschungsgemeinschaft in the 
Sonderforschungsbereich/Transregio
SFB/TR-9 ``Computational Particle Physics'',  by INTAS (grant
03-51-4007) and by RFBR (grant 05-02-17645).

\end{document}